\documentclass[conference]{IEEEtran}
\IEEEoverridecommandlockouts

\usepackage{amsmath,amssymb,amsfonts}
\usepackage{mathtools}
\usepackage{cases}
\usepackage[T1]{fontenc}
\usepackage{textcomp}

\usepackage{graphicx}
\usepackage{epstopdf}
\usepackage{epsfig,scalefnt,multirow}
\usepackage{tabularx}
\usepackage{booktabs}
\usepackage[caption=false]{subfig}
\usepackage{stfloats}

\usepackage{algorithm}
\usepackage{algpseudocode}
\usepackage{algpascal}

\usepackage{xcolor}
\usepackage{soul}
\usepackage{relsize}
\usepackage{url}
\usepackage{cite}
\usepackage{ifthen}

\sethlcolor{green}

\makeatletter
\def\endthebibliography{%
 \def\@noitemerr{\@latex@warning{Empty `thebibliography' environment}}%
 \endlist
}
\catcode`\^^I=9
\makeatother

  \def\cC{{\mathcal{C}}}

 \def\cN{{\mathcal{N}}}

\def\argmax{\mathop{\mathrm{argmax}}}

\def\bSigma{{\pmb{\Sigma}}} 
\def\bPhi{{\pmb{\Phi}}}

\def\b0{{\pmb{0}}} 

\def\ba{{\mathbf{a}}}

 \def\bn{{\mathbf{n}}}  
  \def\bs{{\mathbf{s}}} 
 \def\bv{{\mathbf{v}}}  \def\bx{{\mathbf{x}}}
\def\by{{\mathbf{y}}}   

\def\bA{{\mathbf{A}}}   
 \def\bF{{\mathbf{F}}}  \def\bH{{\mathbf{H}}}
\def\bI{{\mathbf{I}}}   
 \def\bN{{\mathbf{N}}}  \def\bP{{\mathbf{P}}}
 \def\bR{{\mathbf{R}}} \def\bS{{\mathbf{S}}} 
  \def\bW{{\mathbf{W}}} 
\def\bY{{\mathbf{Y}}} 

\DeclarePairedDelimiter\norm{\lVert}{\rVert}
\def\BibTeX{{\rm B\kern-.05em{\sc i\kern-.025em b}\kern-.08em
  T\kern-.1667em\lower.7ex\hbox{E}\kern-.125emX}}

\begin{document}
\title{Joint Channel Estimation and Beamforming for Reconfigurable Intelligent Surface Aided MIMO Systems: Sparsity-Based Approach}

\author{
\IEEEauthorblockN{Sung Hyuck Hong$^{1}$ and Junil Choi$^{2}$}
\IEEEauthorblockA{$^{1}$Algorithm Team, FuriosaAI, Inc., Seoul, Republic of Korea, 06036\\$^{2}$School of Electrical Engineering, Korea Advanced Institute of Science and Technology, Daejeon, Republic of Korea, 34141\\Email: sunghyuck.hong@furiosa.ai, junil@kaist.ac.kr}
}

\maketitle
\footnote{This work was done when Sung Hyuck Hong was affiliated with Korea Advanced Institute of Science and Technology. Current affiliation: FuriosaAI.}

\begin{abstract}
Continuous efforts have been devoted to integrate millimeter wave (mmWave) and terahertz (THz) bands into future communication standards in order to overcome the bandwidth shortage problem and achieve high data rates, primarily through developing accompanying technologies that can overcome the severe propagation loss and blockage associated with increased carrier frequency. One of the most notable accompanying technologies is reconfigurable intelligent surface (RIS), which uses a large number of low-cost passive reflecting elements to reconfigure the propagation environments for improved communication performance and coverage. Despite its numerous benefits, RIS can make channel estimation more difficult due to its lack of radio frequency (RF) chains that can perform baseband signal processing. In addition, the cascaded channel structure of RIS-aided communication systems, which differs from that in conventional systems, brings about significant challenges in both channel estimation and beamforming. In this paper, we propose the joint channel estimation and beamforming optimization algorithm for RIS-aided multiple-input multiple-output (MIMO) communication systems. By carefully exploiting the angular sparsity of mmWave/THz channels, our proposed algorithm successfully designs the RIS matrices that not only facilitate the channel estimation process but also achieve the passive beamforming gain through increased channel capacity. Simulation results demonstrate that our proposed algorithm provides the systems of interest with significant improvement in spectral efficiency.
\end{abstract}


\section{Introduction}
Recent wireless communication systems employ relatively high frequency bands in order to meet the ever-increasing mobile traffic demand and resolve the issue of bandwidth shortage \cite{PI11}. Thanks to its potential of achieving extremely high data rates and enabling efficient packing of large antenna elements into small physical dimension, millimeter wave (mmWave) communication has been adopted as a pivotal component of the fifth generation (5G) wireless network \cite{WANG18}. The trend of exploiting high frequency bands is expected to become more apparent with the arise of terahertz (THz) communication, which offers up to hundreds of gigahertz (GHz) of contiguous bandwidth and thus is expected to bring significant improvement in the performance of future wireless communication systems \cite{Akyildiz22}. Both mmWave and THz channels exhibit high sparsity in the angular domain as a result of the reduced diffraction and increased specular propagation, thereby inspiring numerous research efforts on communication systems with inherent channel sparsity \cite{ANDREWS14, Akyildiz20}.

In order to successfully realize the beamforming gains in communication systems with multiple antennas, channel matrices must be accurately estimated prior to precoder and combiner design \cite{UWAECHIA20}. However, channel state information (CSI) acquisition would become even more challenging in future communication systems that are expected to employ a large number of antennas to overcome the severe propagation loss of mmWave and THz carrier signals \cite{ZHAO20}. While reconfigurable intelligent surface (RIS) has also been proposed as an effective technology to harness the benefits of mmWave/THz communications, it can further complicate the channel estimation process not only because the surface lacks sensing capabilities to explicitly obtain the channel matrices, but also because the channel structure in RIS-aided systems differs from that in conventional systems \cite{WU24}. These challenges in channel estimation inevitably degrades the performance of beamforming optimization, which requires accurately estimated channels to successfully achieve its objective of maximizing the spectral efficiency. 

Motivated by these facts, we propose in this paper the joint channel estimation and beamforming algorithm for RIS-aided multiple-input multiple-output (MIMO) systems. Our proposed algorithm achieves significant improvement in spectral efficiency through the judicious use of angular sparsity present in mmWave/THz channels. Even in the absence of any iterative optimization procedure and perfect CSI, our proposed algorithm successfully constructs the pilot beamformers that not only simplify the cascaded channel structure of RIS-aided MIMO systems, but also realize the substantial performance gains that RIS is expected to bring about. By overcoming the limitations of prior works that either (i) require the knowledge of perfect CSI to design high-performing beamformers \cite{BAHINGAYI22, ZHANG25, ZHANG20}  or (ii) fail to exploit the beamforming capabilities of RIS during the channel estimation process \cite{LIU24,LIN22, MASOOD23}, our proposed algorithm is expected to play a pivotal role in enhancing the performance of RIS-aided mmWave/THz MIMO systems.

The remainder of this paper is organized as follows. The system model is described in Section \ref{system_model_ai}. The joint approach to channel estimation and beamforming optimization in RIS-aided systems is described in Section \ref{prob_for}. The proposed sparsity-based algorithm is explained in Section \ref{proposed_alg}. Simulation results and conclusions are presented in Section \ref{sim_result} and Section \ref{conclusion}, respectively.

\textbf{Notation:} Vectors and matrices are represented by lower and upper boldface letters. The transpose, conjugate transpose, inverse, determinant, and rank of the matrix $\bA$ are represented by $\bA^{\mathrm{T}}$, $\bA^{\mathrm{H}}$, $\bA^{-1}$, $\text{det}(\bA)$, and $\text{rank}(\bA)$, respectively.  The Frobenius norm of $\bA$ is denoted by $\norm{\bA}_{\mathrm{F}}$, while $\text{row}(\bA)$ and $\text{col}(\bA)$ indicate the number of rows and columns of $\bA$, respectively. The element in the $i$-th row and $j$-th column of $\bA$ is denoted by $[\bA]_{i,j}$, while $\text{diag}(\bx)$ represents the diagonal matrix that contains the elements of $\bx$ on its main diagonal. The $N \times N$ identity matrix is represented by $\bI_N$, and the set of all $m \times n$ complex matrices is denoted by ${\mathbb{C}}^{m \times n}$. The expectation operator is written as $\mathbb{E}[\cdot]$, while $|{\cdot}|$ represents the absolute value of a scalar. The function $\text{max}(a,0)$ defined for a real number $a$ is expressed as $(a)^{+}$. The function that returns the arguments corresponding to the top-$k$ greatest values is denoted by $\arg\max k\hspace{3pt}(\cdot)$, where $k$ is a positive integer. Finally, the complex and real normal distributions with mean $m$ and variance $\sigma^2$ are respectively denoted by $\cC \cN(m,\sigma^2)$ and $ \cN(m,\sigma^2)$.

\section{System Model} \label{system_model_ai}
\begin{figure}[!t]
\centering
\includegraphics[width=0.9\columnwidth]{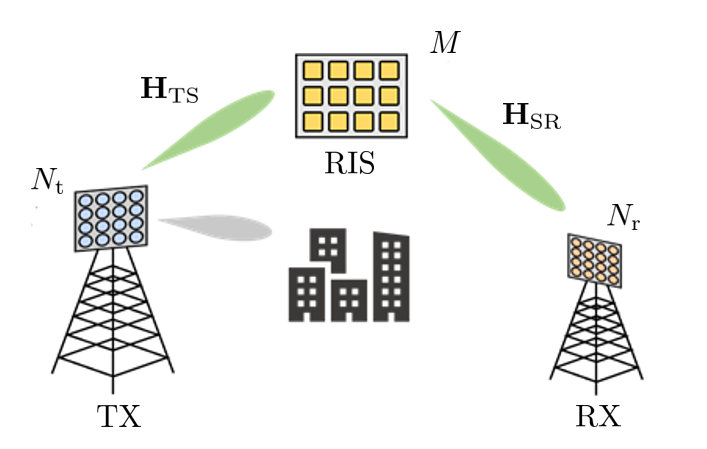}
\caption{Block diagram of an RIS-aided MIMO system. The RIS enables the communication between the TX and RX despite the blockage of the direct link (represented by the gray signal) between them.  }\label{system_diagram}
\end{figure}

This paper studies an RIS-aided MIMO system in which a transmitter (TX) with $N_\text{t}$ antennas communicates with a receiver (RX) with $N_\text{r}$ antennas, as shown in Fig. \ref{system_diagram}. We assume that the direct link between the TX and RX is blocked, a phenomenon that is expected to commonly arise in future communication systems that utilize higher frequency bands \cite{Akyildiz20}. The RIS that consists of $M$ passive reflecting elements facilitates the communication between the TX and RX. The channel from the TX to RIS and that from the RIS to RX are each denoted by $\bH_\text{TS} \in {\mathbb{C}}^{M \times N_\text{t}}$ and $\bH_\text{SR} \in {\mathbb{C}}^{N_\text{r} \times M}$. The total cascaded channel from the TX to RX can then be expressed as 
\begin{align}
\bH_\text{tot}=\bH_\text{SR}\bPhi\bH_\text{TS},\label{cascaded_channel}
\end{align}
where $\bPhi=\text{diag}([e^{j\theta_1},\dots,e^{j\theta_M}]) \in \mathbb{C}^{M \times M}$ and $\theta_m \in [0,2\pi)$ each denote the RIS matrix and the phase shift of the $m$-th RIS element, $m \in \{1,\dots,M\}$. We consider the system of interest to operate in time-division duplexing (TDD) mode, even though the formulation in the paper can be readily extended to frequency-division duplexing (FDD) mode by assuming the use of feedback channel \cite{XIE17_2}.

We primarily consider in this paper mmWave and THz frequency bands, both of which are projected to play a pivotal role in future communication standards \cite{Hemadeh18,Han15}. To accurately describe mmWave/THz channels, we choose the commonly adopted Saleh-Valenzuela model 
\cite{YU16,LIN19,HANSurvey,Lu22} that expresses each of the channel matrices as  
\begin{align}
\bH_i=\sum_{q=0}^{N_\text{path}^i-1}\alpha_{i,q}\ba_\text{r}(\phi^\text{r}_{i,q},\theta^\text{r}_{i,q})\ba_\text{t}(\phi^\text{t}_{i,q},\theta^\text{t}_{i,q})^{\mathrm{H}},\label{channel_matrices_ai}
\end{align} 
where $i \in \{\text{TS, SR}\}$ and $N_\text{path}^i$ respectively denote the channel subscript and the number of propagation paths in $\bH_i$. The complex gain $\alpha_{i,q}$ of the $q$-th path in $\bH_i$ is independently distributed with $\cC\cN(0,\gamma_i^2 10^{-0.1PL(d_i)}), \forall q \in \{0,\dots,N_\text{path}^i-1\}$, where   $\gamma_i=\sqrt{\text{row}(\bH_i)\text{col}(\bH_i)/N_\text{path}^i}$ represents the normalization factor associated with $\bH_i$. Without loss of generality, we assume that $|\alpha_{i,m}| \geq |\alpha_{i,n}|, \forall m, n \in \{0,\dots, N_\text{path}^{i}-1\}$ such that $m<n$. As will be explained in Section \ref{sim_result}, the distance-dependent path loss model $PL(d_i)$ is appropriately adjusted according to the frequency band of interest. The distance between the TX and RIS and that between the RIS and RX are respectively given by $d_\text{TS}$ and $d_\text{SR}$. The azimuth (elevation) angles of arrival and departure (AoAs and AoDs) of the $q$-th path in $\bH_i$ are each given by $\phi^\text{r}_{i,q} (\theta^\text{r}_{i,q}$) and $\phi^\text{t}_{i,q} (\theta^\text{t}_{i,q})$, with the corresponding receive and transmit array response vectors denoted as $\ba_\text{r}(\phi^\text{r}_{i,q},\theta^\text{r}_{i,q}) \in \mathbb{C}^{\text{row}(\bH_i) \times 1}$ and $\ba_\text{t}(\phi^\text{t}_{i,q},\theta^\text{t}_{i,q}) \in \mathbb{C}^{\text{col}(\bH_i) \times 1}$. For specificity, we assume that uniform planar arrays (UPAs) are installed at the TX, RX, and RIS. The transmit array response vector $\ba_\text{t}(\phi^\text{t}_{\text{SR},q},\theta^\text{t}_{\text{SR},q}) \in \mathbb{C}^{M \times 1}$ of the $q$-th path in $\bH_\text{SR}$ is then given by
\begin{align}
k&\ba_\text{t}(\phi^\text{t}_{\text{SR},q},\theta^\text{t}_{\text{SR},q})\notag\\
&\hspace{5pt}=\frac{1}{\sqrt{M}}\Big[
1,\dots,e^{j\frac{2\pi d}{\lambda}(i_h\sin(\phi^\text{t}_{\text{SR},q})\sin(\theta^\text{t}_{\text{SR},q})+i_v\cos(\theta^\text{t}_{\text{SR},q}))},\notag\\
&\hspace{5pt}\dots,e^{j\frac{2\pi d}{\lambda}((M^h_\text{t}-1)\sin(\phi^\text{t}_{\text{SR},q})\sin(\theta^\text{t}_{\text{SR},q})+(M^v_\text{t}-1)\cos(\theta^\text{t}_{\text{SR},q}))} \Big]^\mathrm{T}, \label{transmit_array_example}
\end{align}
where $\lambda$ and $d$ each denote the signal wavelength and the spacing between the antennas or RIS elements. The horizontal and vertical indexes for the RIS reflecting elements are given by $0\leq i_h < M^h$ and $0\leq i_v < M^v$, where $M=M^hM^v$. We omit the explicit definitions of the other array response vectors for the sake of brevity. 

The TX uses the precoder $\bF \in {\mathbb{C}}^{N_\text{t} \times N_\text{s}}$ to send $N_\text{s}$ data streams to the RX, which processes the received signal with the combiner $\bW \in {\mathbb{C}}^{N_\text{r} \times N_\text{s}}$. The processed received signal is given by
\begin{align}
\by&=\bW^{\mathrm{H}}\bH_\text{tot}\bF\bs + \bW^{\mathrm{H}}\bn, \label{processed_received_signal_ai}
\end{align}  
where the symbol vector $\bs \in {\mathbb{C}}^{N_\text{s} \times 1}$ is defined to satisfy $\mathbb{E}[\bs\bs^\mathrm{H}]=\bI_{N_\text{s}}$, and the precoder $\bF$ meets the total power constraint, i.e., $\norm{\bF}_\mathrm{F}^2 \leq P_\text{TX}$. The entries of an additive white Gaussian noise (AWGN) vector $\bn \in {\mathbb{C}}^{N_\text{r} \times 1}$ are independently and identically distributed (i.i.d) with $\cC \cN(0,\sigma^2_{n})$. While we do not explicitly include hybrid beamforming architecture in this paper, the system model of interest can be extended to such scenario thanks to the various prior works on hybrid beamformer design \cite{Kutty16}. The achievable spectral efficiency with the normally distributed transmitted symbols is written as 
\begin{align}
R&=\log_2\text{det}(\bI_{N_\text{s}}+\bR_{\bar{\bn}}^{-1}\bW^\mathrm{H}\bH_\text{tot}\bF\bF^\mathrm{H}\bH_\text{tot}^\mathrm{H} \bW),
\label{spectral_efficiency1}
\end{align}
where $\bR_{\bar{\bn}}=\sigma^2_{n}\bW^\mathrm{H}\bW$ denotes the covariance matrix of the noise $\bar{\bn}=\bW^{\mathrm{H}}\bn$ after combining.

\section{Joint Approach to Channel Estimation and Beamforming Optimization}\label{prob_for}
In this section, we first formulate the problem of channel estimation and beamforming optimization in RIS-aided MIMO systems. We then use the Bayesian framework to explain the motivation and contribution of our proposed algorithm. 

\subsection{Problem Formulation}
The problem of beamforming optimization in RIS-aided MIMO systems can be written as 
\begin{subequations}
\begin{align}
&\text{(P1)}\hspace{50pt} \max_{\bW,\bPhi,\bF} \medspace R \\ 
&\hspace{4pt}\text{subject to } \hspace{7pt}\bPhi=\text{diag}([e^{j\theta_1},\dots,e^{j\theta_M}]), \\
&\hspace{54pt}\bR_{\bar{\bn}}=\sigma^2_{n}\bW^\mathrm{H}\bW,\\
&\hspace{54pt}\bH_\text{tot}=\bH_\text{SR}\bPhi\bH_\text{TS},\\
&\hspace{54pt}\norm{\bF}_\mathrm{F}^2 \leq P_\text{TX}.
\end{align}
\end{subequations}
In other words, it aims to maximize the spectral efficiency by designing the beamformer $\bA^\star = \{\bW^\star,\bPhi^\star,\bF^\star\}$ suited to the given channel realizations $\bH_\text{TS}$ and $\bH_\text{SR}$.
However, $\bH_\text{TS}$ and $\bH_\text{SR}$ are usually not known and thus must be estimated prior to constructing beamformers.

To estimate the channel matrices in practical communication systems, the RX processes the pilot signals that are transmitted by the TX. The processed received pilot signal at the time index $t \in \{0,\dots,T-1\}$ is expressed as 
\begin{align}
\by_p[t]=\bW^\mathrm{H}_p[t]\bH_\text{SR}\bPhi_p[t]\bH_\text{TS}\bF_p[t]\bs[t]+\bW^\mathrm{H}_p[t]\bn[t], \label{recevied_signal_ai}
\end{align} 
where $\bP[t]=\{\bW_p[t], \bPhi_p[t], \bF_p[t]\}$ is the pilot beamformer, $\bs[t]$ is the symbol vector, and each element of $\bn[t]$ is i.i.d with $\cC \cN(0,\sigma^2_{n})$. The pilot beamformers $\bP=\{\bP[0],\dots,\bP[T-1]\}$ must be constructed so that $\bH_\text{TS}$ and $\bH_\text{SR}$ can be accurately estimated from the received pilot signal matrix $\bY_p = [\by_p[0],\dots,\by_p[T-1]]$. Note that the cascaded channel structure in \eqref{cascaded_channel} sets RIS-aided systems apart from conventional counterparts and necessitates novel estimation methods \cite{WU24}.

\subsection{Bayesian Perspective on Proposed Algorithm}
\begin{figure}[!t]
\centering
\includegraphics[width=0.7\columnwidth]{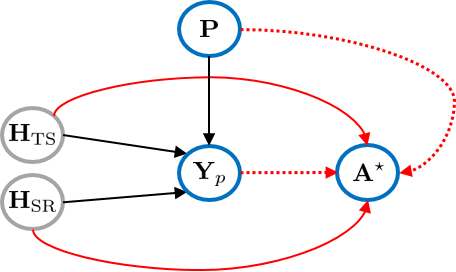}
\caption{Bayesian network of the joint channel estimation and beamforming optimization problem in RIS-aided MIMO system. Gray (blue) circles represent the latent (observable) variables, while black (red) arrows indicate the conditional dependencies associated with channel estimation (beamforming optimization). Unlike previous works, our algorithm directly exploits the red dotted arrows by incorporating $\bP$ and $\bY_p$ into the design of $\bA^\star$.}   \label{bayesian_network}
\end{figure}

Practical communication systems without CSI usually perform channel estimation and beamforming optimization in a sequential manner. Joint channel estimation and beamforming optimization is thus associated with the posterior distribution 
\begin{align}
&p(\bA^\star,\bH_\text{TS},\bH_\text{SR} | \bY_p,\bP)\notag\\&\hspace{40pt}=p(\bH_\text{TS},\bH_\text{SR} |\bY_p,\bP)p(\bA^\star| \bY_p,\bP, \bH_\text{TS},\bH_\text{SR}),\label{bayesian_eq}
\end{align}
where $p(\bH_\text{TS},\bH_\text{SR} |\bY_p, \bP)$ and $p(\bA^\star| \bY_p,\bP, \bH_\text{TS},\bH_\text{SR})$ respectively describe the posterior distributions of interest during the pilot-aided channel estimation and beamforming optimization. Fig. \ref{bayesian_network} shows the Bayesian network of the joint channel estimation and beamforming optimization problem in RIS-aided MIMO system. The latent variables $\bH_\text{TS}$ and $\bH_\text{SR}$ are inferred from the observable variables $\bY_p$ and $\bP$ during the channel estimation, while the beamforming optimization involves constructing the beamformer $\bA^\star$ that maximizes the spectral efficiency $R$ in \eqref{spectral_efficiency1}. The red dotted arrows from $\bP$ to $\bA^\star$ and from $\bY_p$ to $\bA^\star$ show that 
the pilot beamformers $\bP$, if carefully designed, can also be effectively utilized during the beamforming optimization process. However, this opportunity was  overlooked in prior works that limited the use of $\bP$ and $\bY_p$ to estimate channels, i.e., they adopted the approximation 
\begin{align}
p(\bA^\star| \bY_p,\bP, \bH_\text{TS},\bH_\text{SR}) \approx p(\bA^\star| \bH_\text{TS},\bH_\text{SR})
\end{align}
for the sake of simplicity but at the cost of performance degradation \cite{LIU24,LIN22, MASOOD23}. 

Motivated by these facts, we propose the joint channel estimation and beamforming optimization algorithm for RIS-aided MIMO systems. In addition to accurately estimating the channel matrices, the proposed algorithm successfully reaps the considerable capacity gains from RIS through the judicious processing of received signals. Unlike prior works, our proposed algorithm directly leverages the pilot signals for passive beamforming optimization and thus is highly suitable for future communication systems that aim to support extremely high spectral efficiency and low latency at the same time \cite{Almekhlafi21}.

\section{Proposed Joint Channel Estimation and Beamforming Optimization Algorithm} \label{proposed_alg}
As explained in Section \ref{prob_for}, it can be highly beneficial to exploit the pilot beamformers and received pilot signals to not only estimate the channel matrices but also optimize the beamformers in RIS-aided MIMO systems. Motivated by this, we propose in this section the joint channel estimation and beamforming optimization algorithm that achieves significant improvement in spectral efficiency. In contrast to the majority of the beamforming methods that assume the availability of perfect CSI, the proposed algorithm exploits the channel sparsity present in the systems of interest so as to efficiently obtain the beamformer $\bA^\star = \{\bW^\star, \bPhi^\star, \bF^\star\}$ from the received pilot signal matrix $\bY_p$ and AoA/AoD parameters, the latter of which remain relatively constant during the coherence time and thus can be acquired with the aid of various proposed estimation techniques \cite{ZHANG19}.

Let us respectively denote the path indexes for $\bH_\text{TS}$ and $\bH_\text{SR}$ as $i \in \{0,...,N_\text{path}^{\text{TS}}-1\}$ and  $j \in \{0,...,N_\text{path}^{\text{SR}}-1\}$. Then, consider setting the RIS pilot matrix as 
\begin{align}
\bPhi_p[t]=\bPhi^{i,j}_p=\text{diag}(\bv_p^{i,j}),\label{Phi_i,j} 
\end{align} 
where $\bv_p^{i,j}=M\text{diag}(\ba_\text{r}(\phi^\text{r}_{\text{TS},i},\theta^\text{r}_{\text{TS},i}))^\mathrm{H}\ba_\text{t}(\phi^\text{t}_{\text{SR},j},\theta^\text{t}_{\text{SR},j})$ is the scaled UPA array response vector constructed with the $i$-th receive path of $\bH_\text{TS}$ and $j$-th transmit path of $\bH_\text{SR}$. Without the loss of generality, we omit the time index $t$ throughout the rest of this section for the sake of notational convenience. The asymptotic orthogonality of UPA array response vectors ensures that
\begin{align}
\bH_\text{SR}\bPhi_p^{i,j}\bH_\text{TS}\rightarrow \alpha_{\text{TS},i}\alpha_{\text{SR},j}\ba_\text{r}(\phi^\text{r}_{\text{SR},j},\theta^\text{r}_{\text{SR},j})\ba_\text{t}(\phi^\text{t}_{\text{TS},i},\theta^\text{t}_{\text{TS},i})^\mathrm{H}\label{H_tot_i,j},  
\end{align}
as $M \rightarrow \infty$ \cite{HONG22}. Since the systems of interest are expected to deploy a large number of passive reflecting elements, setting the RIS matrix as $\bPhi_p^{i,j}$ in \eqref{Phi_i,j} has the effect of simplifying the cascaded structure of  $\bH_\text{tot}=\bH_\text{SR}\bPhi\bH_\text{TS}$ and thus enables the accurate channel estimation despite the absence of baseband signal processing capabilities at the RIS \cite{Zhang21}. As will be shown later, $\bPhi^{i,j}_p$ in \eqref{Phi_i,j} offers the additional benefit of increasing the capacity of $\bH_\text{tot}$, thereby successfully reconfiguring the propagation environments for improved communication.

By defining the symbol matrices as $\bS_p = [\bs[0],\ldots \bs[N_\text{s}-1]] \in {\mathbb{C}}^{N_\text{s} \times N_\text{s}}$, we can express the received pilot signal matrix $\bY_p^{i,j}$ corresponding to the adjusted total channel $\bH_{\text{tot},(i,j)}=\bH_\text{SR}\bPhi^{i,j}_p\bH_\text{TS}$ as 
\begin{align}
\bY^{i,j}_p&=\bW_p^\mathrm{H}\bH_\text{SR}\bPhi^{i,j}_p\bH_\text{TS}\bF_p\bS_p+\bW_p^\mathrm{H}\bN^{i,j} \notag \\ 
&=\bW_p^\mathrm{H}\bH_{\text{tot},(i,j)}\bF_p\bS_p+\bW_p^\mathrm{H}\bN^{i,j}\label{Y_sup_i,j},
\end{align}
where each column vector of $\bN^{i,j}$ is identically distributed with $\bn[t]$ in \eqref{recevied_signal_ai}. In order to accurately estimate $\bH_{\text{tot},(i,j)}$ by exploiting its asymptotic property described in \eqref{H_tot_i,j}, we set the pilot combiner $\bW_p\in{\mathbb{C}}^{N_\text{r} \times N_\text{s}}$ and precoder $\bF_p\in{\mathbb{C}}^{N_\text{t} \times N_\text{s}}$ as
\begin{align}
\bW_\text{p}&=\begin{bmatrix}
\ba_\text{r}(\phi^\text{r}_{\text{SR},0},\theta^\text{r}_{\text{SR},0}) & \cdots & \ba_\text{r}(\phi^\text{r}_{\text{SR},N_\text{s}-1},\theta^\text{r}_{\text{SR},N_\text{s}-1})   
\end{bmatrix}, \\
\bF_\text{p}&=\begin{bmatrix}
\ba_\text{t}(\phi^\text{t}_{\text{TS},0},\theta^\text{t}_{\text{TS},0}) & \cdots & \ba_\text{t}(\phi^\text{t}_{\text{TS},N_\text{s}-1},\theta^\text{t}_{\text{TS},N_\text{s}-1})   
\end{bmatrix}.
\end{align}
It then holds that the element $\max(\bW_p^\mathrm{H}\bH_{\text{tot},(i,j)}\bF_p)$ of $\bW_p^\mathrm{H}\bH_{\text{tot},(i,j)}\bF_p$ with the maximum absolute value satisfies 
\begin{align}
\text{max}(\bW_p^\mathrm{H}\bH_{\text{tot},(i,j)}\bF_p)
&\rightarrow \alpha_{\text{TS},i}\alpha_{\text{SR},j}\label{asy_Y_sup_i,j},
\end{align}
as $N_\text{r}, N_\text{t}, M \rightarrow \infty$. As a result, the combined complex gain $\alpha_{\text{TS},i}\alpha_{\text{SR},j}$ can be estimated by 
$[\tilde{\bY}^{i,j}_p ]_{r^{i,j}_0,c^{i,j}_0}$, where $\tilde{\bY}^{i,j}_p=\bY^{i,j}_p\bS_p^{-1}$ is the equalized received signal matrix and
\begin{align}
\{r^{i,j}_z,c^{i,j}_z\}_{z=0,...,k-1}= \argmax_{r,c
} k\hspace{3pt} \left(|[\tilde{\bY}^{i,j}_p]_{r,c}|\right).
\end{align}
The indexes $r_z^{i,j}$ and $c_z^{i,j}$ correspond to the row and column indexes of the element of $\tilde{\bY}^{i,j}_p$ with the top-$(z+1)$ absolute value, and the hyperparameter $k \in \{1,\dots, N_\text{s}\}$ defines the number of elements to sample from $\tilde{\bY}^{i,j}_p$. The result in \eqref{asy_Y_sup_i,j} ensures that  $[\tilde{\bY}^{i,j}_p]_{r^{i,j}_0,c^{i,j}_0}$ approaches the unbiased estimator of $\alpha_{\text{TS},i}\alpha_{\text{SR},j}$ in the limit of $N_\text{r}, N_\text{t}, M$, demonstrating the suitability of the proposed estimator for RIS-aided MIMO systems with channel sparsity. The rank-$N_\text{rank}$
approximation of $\bH_{\text{tot},(i,j)}$ can then be written as  
\begin{align}
&\hat{\bH}^{N_\text{rank}}_{\text{tot},(i,j)}=\notag\\
&\sum_{z=0}^{N_\text{rank}-1} [\tilde{\bY}^{i,j}_p]_{r^{i,j}_z,c^{i,j}_z}\ba_\text{r}(\phi^\text{r}_{\text{SR},r^{i,j}_z},\theta^\text{r}_{\text{SR},r^{i,j}_z})\ba_\text{t}(\phi^\text{t}_{\text{TS},c^{i,j}_z},\theta^\text{t}_{\text{TS},c^{i,j}_z})^\mathrm{H}, \label{rank_approx}
\end{align}
where $N_\text{rank} \in \{1,...,k\}$. As each of $\bW_p$ and $\bF_p$ becomes semi-unitary in the limit of its row dimension, $\hat{\bH}^{N_\text{rank}}_{\text{tot},(i,j)}$ in \eqref{rank_approx} can be considered as the approximation of the truncated singular value decomposition (SVD) of $\bH_\text{SR}\bPhi_p^{i,j}\bH_\text{TS}$, whose angular sparsity allows our proposed algorithm to efficiently acquire CSI without resorting to any iterative method \cite{Li18}.
 
We now leverage the proposed complex gain estimator  to realize the spectral efficiency gain that RIS is expected to bring about. As mmWave and THz channels typically exhibit the presence of one dominant path pair, we set the RIS matrix as $\bPhi^\star =\bPhi_p^{i^\star,j^\star}$, where 
\begin{align}
i^\star, j^\star = 
\argmax_{i,j
} 1\hspace{3pt} \left({|[\tilde{\bY}^{i,j}_p]_{r^{i,j}_0,c^{i,j}_0}|}\right). \label{choice}
\end{align}
In other words, we use the complex gain estimator $\{[\tilde{\bY}^{i,j}_p]_{r^{i,j}_0,c^{i,j}_0}\}_{\forall i,j}$ to determine $\bH_{\text{tot},(i^\star,j^\star)}$ that is likely to achieve the highest capacity among $\{\bH_{\text{tot},(i,j)}\}_{\forall i,j}$. In fact,  the chosen path pair in \eqref{choice} is guaranteed to be asymptotically optimal in the limit of large arrays and vanishing noise, i.e., 
\begin{align}
C(\bH_{\text{tot},(i^\star,j^\star)}) \rightarrow \max_{i,j}C(\bH_{\text{tot},(i,j)}),\label{asy_capacity}
\end{align}  
as $N_\text{r}, N_\text{t}, M \rightarrow \infty$ and  $\sigma_n \rightarrow 0$. In \eqref{asy_capacity}, $C(\bH)$ denotes the capacity of the channel matrix $\bH$ of appropriate dimension. Finally, the corresponding combiner and precoder are obtained as 
\begin{align}
\bW^\star&=\bW_\text{opt}(\hat{\bH}^{N_\text{rank}}_{\text{tot},(i^\star,j^\star)}, N_\text{s}),\\\bF^\star&=\bF_\text{opt}(\hat{\bH}^{N_\text{rank}}_{\text{tot},(i^\star,j^\star)}, N_\text{s}, P_\text{TX}),
\end{align}
where the optimal combiner $\bW_\text{opt}(\bH, N_\text{s})$ and precoder $\bF_\text{opt}(\bH, N_\text{s}, P_\text{TX})$ for $\bH$ under the transmit power constraint $P_\text{TX}$ can be obtained by SVD and waterfilling. We finally note that the effect of imperfect orthogonality of array response vectors in the finite antenna regime can be compensated by setting $N_\text{rank}>1$ for $\hat{\bH}^{N_\text{rank}}_{\text{tot},(i,j)}$ in \eqref{rank_approx}. 

\begin{figure}[!t]
\centering
\includegraphics[width=0.9\columnwidth]{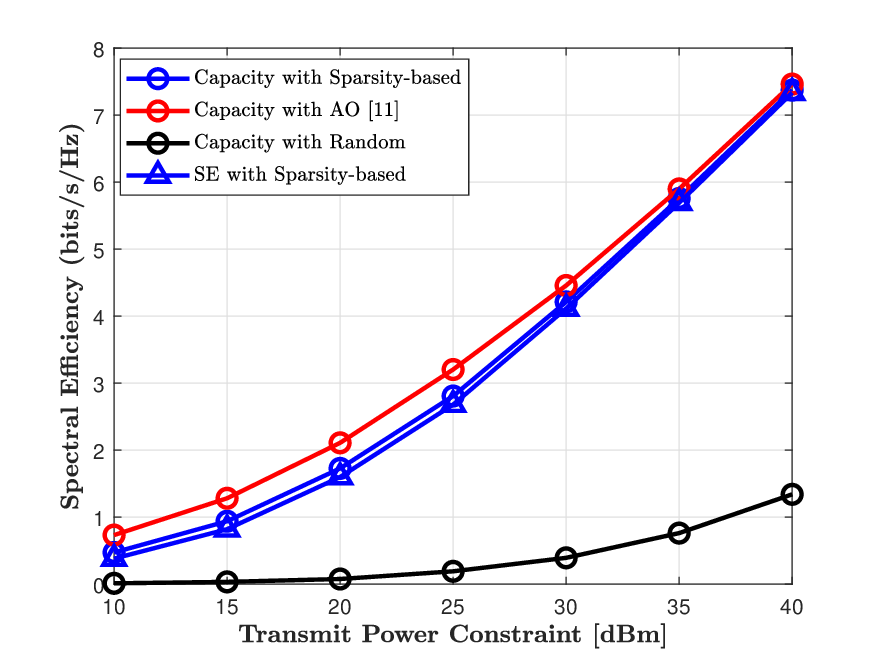}
\caption{Capacity and spectral efficiency achieved in RIS-aided mmWave MIMO systems versus the transmit power constraint $P_\text{TX}$.}   \label{ai_sim}
\end{figure}
\begin{figure}[!t]
\centering
\includegraphics[width=0.9\columnwidth]{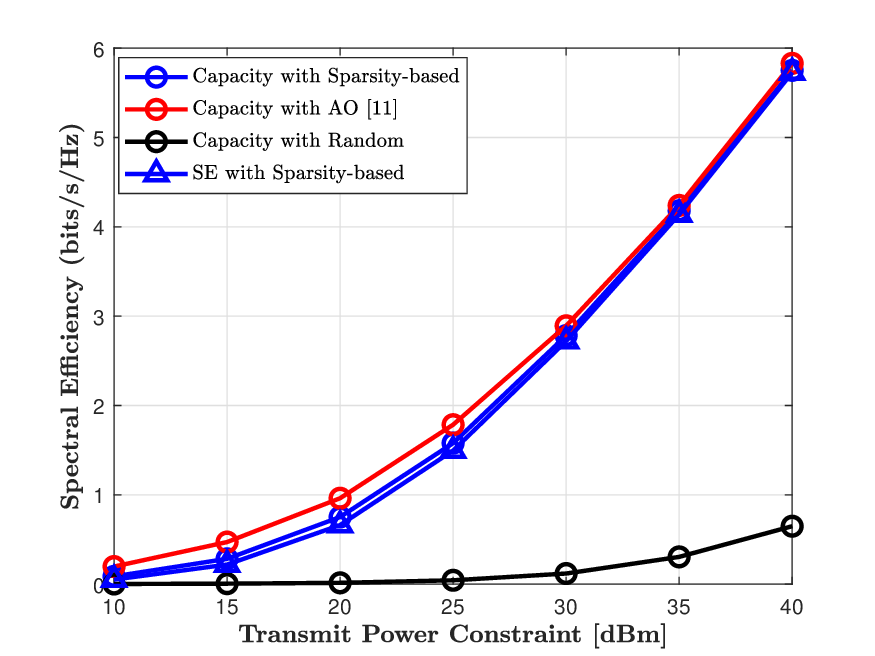}
\caption{Capacity and spectral efficiency achieved in RIS-aided THz MIMO systems versus the transmit power constraint $P_\text{TX}$.}   \label{ai_sim_2}
\end{figure}
\section{Simulation Results} \label{sim_result}
In this section, we present simulation results to demonstrate the effectiveness of our proposed channel estimation and beamforming optimization algorithm. We assume that both the TX and RX are equipped with $N_\text{t}=N_\text{r}=8 \times 8=64$ antennas, while the RIS is equipped with $M=16 \times 16 =256$ passive reflecting elements. The number of paths is set to be $N_\text{path}^{\text{TS}}=N_\text{path}^{\text{SR}}=4$, while AoAs/AoDs are sampled from the uniform distribution. The distance-dependent path loss model $PL(d_i)$ is given by
\begin{align}
PL(d_i)\medspace[\text{dB}]=\alpha+10\beta\log_{10}(d_i)+\xi, \label{PL_model}
\end{align} 
where $\xi\sim\cN(0,\sigma^2)$. We consider both mmWave and THz channels by choosing the appropriate values for the parameters in \eqref{PL_model}. For 28 GHz mmWave channel, the parameters are set to be $\alpha=61.4, \beta=2, \sigma=5.8\medspace\text{dB}$ for a line-of-sight (LOS) path of $\bH_i$, and $\alpha=72.0, \beta=2.92, \sigma=8.7\medspace\text{dB}$ for its non-line-of-sight (NLOS) paths \cite{AKDENIZ14}. For THz channel, we consider the usage of 142 GHz carrier frequency by respectively choosing $\alpha=75.44, \beta=2.1, \sigma=2.8\medspace\text{dB}$ and  $\alpha=75.44, \beta=3.1, \sigma=8.3\medspace\text{dB}$ for LOS and NLOS paths \cite{XING21}. Other simulation parameters are set as follows: $N_\text{rank}=1, N_\text{s}=4, d_\text{TS} = 35 \text{ m}, d_\text{SR} = 15 \text{ m}, d=\lambda/2$, and  $\sigma^2_{n}=-91 \medspace\text{dBm}$. All the simulation results are averaged over 10,000 independent channel realizations. 

Fig. \ref{ai_sim} shows the capacity and spectral efficiency achieved by the proposed sparsity-based algorithm and other benchmarks in RIS-aided mmWave MIMO systems. The capacity $C(\bH_{\text{tot},(i^\star,j^\star)})$ of the sparsity-based algorithm is defined as the maximum spectral efficiency achievable under the total cascaded channel $\bH_{\text{tot},(i^\star,j^\star)}=\bH_\text{SR}\bPhi_p^{i^\star,j^\star}\bH_\text{TS}$ and the transmit power constraint $P_\text{TX}$, i.e, 
\begin{align}
C(\bH_{\text{tot},(i^\star,j^\star)})=\sum_{l=1}^{N_\text{s}}\log_2\left(1+\frac{P_l}{\sigma^2_{n}}|[\bSigma^\star_\text{tot}]_{l,l}|^2\right) \label{derivation_obj_effective_channel}.
\end{align}
In \eqref{derivation_obj_effective_channel}, $\bSigma^\star_\text{tot} \in {\mathbb{C}}^{N_\text{r} \times N_\text{t}}$ contains the singular values of $\bH_{\text{tot},(i^\star,j^\star)}$ on its main diagonal, $P_l=\left(\frac{1}{\hat{\eta}}-\frac{\sigma^2_{n}}{|[\bSigma_\text{tot}]_{l,l}|^2}\right)^+$ represents the power allocated to the $l$-th eigenchannel of $\bH_{\text{tot},(i^\star,j^\star)}$, and $\hat{\eta}$ is determined such that $\sum_{l=1}^{N_\text{s}}P_l=P_\text{TX}$ holds. The capacity of 
the alternating optimization (AO)-based algorithm can be similarly defined as $C(\bH_{\text{tot,AO}})$, where $\bH_{\text{tot,AO}}$ is the total channel adjusted according to the method described in \cite{ZHANG20}. For comparison, we also included the capacity attained when the phase shift of each RIS element is randomly selected from $[0,2\pi)$. The result shows that the RIS matrix constructed by the proposed algorithm performs close to that constructed by the AO-based algorithm, which assumes the availability of perfect CSI and requires the computational complexity of $\mathcal{O}(3MN_\text{r}^3+2MN_\text{r}^2N_\text{t})$. In contrast, by carefully exploiting the channel sparsity in RIS matrix design, our proposed algorithm achieves substantial capacity gains while requiring only the AoA/AoD parameters and significantly reduced complexity of $\mathcal{O}(MN_\text{r}N_\text{t})$. It should also be noted that the relatively poor performance of random RIS design shows that the performance gain from a large number of RIS elements can only be harnessed through sophisticated passive beamforming optimization, further underscoring the importance of 
our proposed algorithm. Furthermore, the small difference between the capacity and spectral efficiency of the sparsity-based algorithm demonstrates its capability of accurately estimating the total cascaded channels of RIS-aided MIMO systems. 

In Fig. \ref{ai_sim_2}, the capacity and spectral efficiency achieved by the proposed sparsity-based algorithm and other benchmarks in RIS-aided THz MIMO systems are plotted as a function of $P_\text{TX}$. The result shows that the proposed algorithm achieves spectral efficiency close to the capacity attained by the AO-based algorithm, thereby demonstrating its suitability for THz channels as well. It can thus be concluded that the proposed joint channel estimation and beamforming optimization algorithm can attain significant spectral efficiency gains in future communication systems, which are expected to deploy RIS in order to fully reap the performance benefits that mmWave and THz communications are envisioned to offer.

\section{Conclusions} \label{conclusion}
In this paper, we proposed the joint channel estimation and beamforming optimization algorithm for RIS-aided MIMO systems. The proposed sparsity-based algorithm successfully constructs the pilot matrices that not only facilitate CSI acquisition process but also provide the systems of interest with considerable capacity gains, thereby resolving the inefficiency involved with the conventional method of limiting the usage of pilot signals to channel estimation. Effectively utilizing the channel sparsity present in future communication systems, our proposed algorithm is capable of achieving significant spectral efficiency improvement while requiring limited computational complexity.

\section{Acknowledgement} \label{acknowledgement}
This work was supported in part by the Institute of Information \& Communications Technology Planning \& Evaluation (IITP)-Information Technology Research Center (ITRC) grant funded by the Ministry of Science and ICT (MSIT) of the Korea government (IITP-2025-RS-2020-II201787, contribution rate 50\%) and in part by IITP under 6G Cloud Research and Education Open Hub grant funded by MSIT. (IITP-2025-RS-2024-00428780, contribution rate 50\%)

\bibliographystyle{ieeetr}
\bibliography{GLOBECOM_refs}
\end{document}